\documentclass[sigconf, screen, nonacm]{acmart}
   
\setcopyright{none}

\begin{document}

\title{Cost-Aware cHBM-Array Accelerator for LLM Decoding}

\author{Daegun Yoon}
\affiliation{%
    \institution{SK hynix}
    \city{Icheon}
    \country{Republic of Korea}
}
\email{daegun.yoon@sk.com}

\author{Yunki Han}
\affiliation{%
    \institution{SK hynix}
    \city{Icheon}
    \country{Republic of Korea}
}
\email{yunki.han@sk.com}

\author{Jinlin Wu}
\affiliation{%
    \institution{Illinois Institute of Technology}
    \city{Chicago}
    \state{IL}
    \country{USA}
}
\email{jwu141@hawk.illinoistech.edu}

\author{Soohong Ahn}
\affiliation{%
    \institution{SK hynix}
    \city{Icheon}
    \country{Republic of Korea}
}
\email{soohong.ahn@sk.com}

\author{Jiwan Kim}
\affiliation{%
    \institution{SK hynix}
    \city{Icheon}
    \country{Republic of Korea}
}
\email{jiwan2.kim@sk.com}

\author{Nameun Kang}
\affiliation{%
    \institution{SK hynix}
    \city{Icheon}
    \country{Republic of Korea}
}
\email{nameun.kang@sk.com}

\author{Myunghyun Rhee}
\affiliation{%
    \institution{SK hynix}
    \city{Icheon}
    \country{Republic of Korea}
}
\email{myunghyun.rhee@sk.com}

\author{Donghyeon Yi}
\affiliation{%
    \institution{SK hynix}
    \city{Icheon}
    \country{Republic of Korea}
}
\email{donghyeon.yi@sk.com}

\author{Mingoo Kang}
\affiliation{%
    \institution{SK hynix}
    \city{Icheon}
    \country{Republic of Korea}
}
\email{michael.kang@sk.com}

\author{Joohyung Kim}
\affiliation{%
    \institution{SK hynix}
    \city{Icheon}
    \country{Republic of Korea}
}
\email{joohyung.kim@sk.com}

\author{Ieryung Park}
\affiliation{%
    \institution{SK hynix}
    \city{Icheon}
    \country{Republic of Korea}
}
\email{ieryung.park@sk.com}

\author{Minxuan Zhou}
\affiliation{%
    \institution{Illinois Institute of Technology}
    \city{Chicago}
    \state{IL}
    \country{USA}
}
\email{mzhou26@illinoistech.edu}

\author{Youngpyo Joo}
\affiliation{%
    \institution{SK hynix}
    \city{Icheon}
    \country{Republic of Korea}
}
\email{youngpyo.joo@sk.com}

\author{Hoshik Kim}
\affiliation{%
    \institution{SK hynix}
    \city{Icheon}
    \country{Republic of Korea}
}
\email{hoshik.kim@sk.com}

\author{Joonseop Sim}
\affiliation{%
    \institution{SK hynix}
    \city{Icheon}
    \country{Republic of Korea}
}
\email{joonseop.sim@sk.com}

\renewcommand{\shortauthors}{Daegun Yoon et al.}

\begin{abstract}
Autoregressive decoding of mixture-of-experts (MoE) large language models is predominantly memory-bound due to low arithmetic intensity of attention and MoE layers.
Conventional GPU--HBM architectures face a GPU shoreline bottleneck limiting aggregate memory bandwidth scaling with more HBM stacks.
While 3D-stacked near-memory architectures can overcome this bottleneck by vertically integrating compute logic with DRAM, existing designs often incur prohibitive area, power, and packaging overheads from overprovisioning compute and die-to-die (D2D) resources.
Resolving these tradeoffs requires an architecture that jointly optimizes performance, physical feasibility, and manufacturing cost.

We propose CASCADE, a cost-aware memory-centric accelerator that executes the entire decode phase on an array of low-power, compute-enabled custom HBM (cHBM) chiplets.
Through workload-driven design space exploration (DSE), CASCADE co-optimizes compute throughput, D2D bandwidth, and HBM stack height against physical layout and cost constraints, enabling low-cost standard organic-substrate packaging.
To efficiently utilize its distributed resources, CASCADE employs two-dimensional parallelism strategies.
Because the optimal strategy for CASCADE varies with workload characteristics, we identify the optimal one for each workload through DSE.
CASCADE achieves up to $4.65\times$ higher decode throughput and $8.26\times$ higher energy efficiency than state-of-the-art accelerators, with an estimated 38.4\% lower per-package manufacturing cost than a prior chiplet-array architecture.
\end{abstract}

{\hypersetup{urlcolor=black}
\maketitle
}

\section{Introduction}\label{sec:1}
Mixture-of-experts (MoE) architectures have emerged as an efficient approach to scaling large language models (LLMs) by activating only a subset of experts for each token~\cite{switchtransformers}. Despite their computational efficiency, MoE-LLM decoding remains challenging, particularly at small batch sizes. Attention repeatedly accesses growing key-value (KV) caches~\cite{flashattention,flashinfer}, while expert routing limits weight reuse in MoE grouped general matrix multiplications (GEMMs)~\cite{sonicmoe}. Unlike dense projections, which benefit from larger batches, MoE experts process only their routed tokens, limiting improvements in arithmetic intensity. Our profiling shows that memory-bound grouped-query attention (GQA) and MoE grouped GEMMs dominate decode time, while linear projections occupy only a small fraction. Consequently, the substantial compute resources of conventional GPUs remain underutilized, and accelerating attention alone leaves another major memory bottleneck unresolved. Addressing both operations efficiently requires overcoming three interrelated architectural challenges.

\noindent\textbf{Challenge 1: Scaling memory bandwidth beyond GPU shoreline.}
Conventional GPU--HBM architectures cannot readily scale aggregate memory bandwidth by adding HBM stacks. Each stack requires a wide interface along the finite GPU die perimeter, known as GPU shoreline~\cite{memory_fabric}. Processing-near-memory (PNM) and processing-in-memory (PIM) architectures leverage 3D stacking to exploit high internal memory bandwidth via vertical links~\cite{hbm_pim,duplex,stratum}. However, increasing bandwidth within a memory stack and scaling aggregate bandwidth across a package address distinct limitations. For example, Duplex~\cite{duplex} combines an xPU with PIM units in HBM logic dies (Logic-PIM), exploiting high internal bandwidth for attention and MoE. Nevertheless, its central xPU still relies on GPU--HBM interfaces, retaining the shoreline constraint on adding more stacks. Furthermore, its fourfold expansion of per-stack internal bandwidth through TSVs requires greater logic-die compute throughput to fully exploit the additional bandwidth~\cite{duopd}. An alternative approach replaces the central GPU with an array of interconnected, compute-enabled HBM chiplets, as demonstrated by AMMA~\cite{amma}. By allowing each chiplet to access its local DRAM, this architecture can scale aggregate bandwidth with stack count rather than GPU shoreline. However, realizing this bandwidth advantage introduces new constraints on chiplet implementation and interconnection.

\noindent\textbf{Challenge 2: Balancing physical feasibility and manufacturing cost.}
Scaling bandwidth through chiplet arrays requires careful allocation of limited hardware resources. HBM logic dies must accommodate TSVs, memory PHYs, and test circuitry~\cite{samsung_hbm4,samsung_chbm4}, limiting the available area for processing units and die-to-die (D2D) interfaces. Increasing compute throughput beyond what memory-bound workloads can effectively utilize wastes valuable die area and power without proportional performance gains. Similarly, D2D interfaces provisioned with excessive bandwidth consume PHY area and routing resources and may require costly advanced packaging~\cite{ucie_packaging,chipletcost}, despite limited decoding performance gains. HBM stack height introduces another tradeoff: additional DRAM core dies increase memory capacity and manufacturing costs~\cite{hbm_packaging} without increasing nominal per-stack bandwidth beyond HBM4 4-Hi~\cite{semianalysis_4hi}. AMMA~\cite{amma} demonstrates the potential of chiplet arrays using high compute throughput and four-directional, high-bandwidth D2D interfaces. However, its hardware specifications are not comprehensively justified through a joint assessment of workload requirements, logic-die area, power, and packaging costs. The key challenge is therefore to determine how much compute throughput, interconnect bandwidth, and memory capacity are actually needed, while ensuring that the resulting architecture is physically feasible and cost-effective.

\noindent\textbf{Challenge 3: Efficient execution under constrained interconnect bandwidth.}
Reducing hardware resources alone does not guarantee an efficient accelerator, because execution strategy affects resource utilization and communication requirements. At the system level, attention--FFN disaggregation (AFD)~\cite{afd,afd_challenges} assigns attention and feed-forward network (FFN) or MoE computation to separate accelerator groups, requiring inter-device activation transfers at every Transformer layer while leaving FFN or MoE constrained by GPU memory bandwidth. Prefill--decode disaggregation (PDD)~\cite{splitwise,distserve} instead enables the entire decode phase on memory-centric accelerators without these transfers. Within a chiplet array, tensor parallelism (TP) exploits distributed memory bandwidth but incurs frequent collective communication~\cite{megatronlm} and may require KV cache replication when the TP degree exceeds the number of KV heads~\cite{vllm_dcp}, whereas pipeline parallelism (PP) can underutilize chiplets at small batch sizes. These tradeoffs intensify when D2D bandwidth is constrained to reduce implementation costs. Simply reducing interconnect bandwidth without adapting execution can introduce communication bottlenecks, offsetting the benefits of cost-aware hardware design. Conversely, communication-efficient parallelism can reduce the bandwidth required to sustain high performance. Thus, hardware specifications and execution strategies must be designed together rather than optimized independently.

We propose CASCADE (\underline{c}hiplet-\underline{a}rray-based \underline{sca}lable \underline{de}coder), a cost-aware accelerator comprising a $4\times4$ array of identical compute-enabled custom HBM (cHBM) chiplets. By replacing the central GPU with locally operating chiplets, CASCADE offers 44 TB/s of aggregate HBM bandwidth and executes both attention and MoE computation on the array. Each chiplet accesses only its own HBM stack, using D2D links to exchange intermediate data. Rather than maximizing compute throughput, CASCADE employs its tensor processing capability according to the arithmetic intensity of target decoding operations. We adopt HBM4 8-Hi stacks to accommodate BF16 model weights and long-context KV caches, balancing memory capacity against stack cost and power.

CASCADE determines compute throughput, D2D bandwidth, and memory capacity through workload-driven design space exploration (DSE), constrained by logic-die area, power, and packaging costs. Our D2D bandwidth sensitivity study shows that reducing bandwidth from 1024 to 128 GB/s increases decode latency by only 2.4--4.2\% for GPT-OSS-120B using TP across batches 1--32 at a 100K context. This finding motivates a design that avoids costly bandwidth overprovisioning. Our floorplan accommodates host I/O and three-directional D2D interfaces on identical chiplets, making standard organic-substrate packaging viable under modeled design constraints. We further evaluate die dimensions, reticle utilization, yield, and package costs rather than treating chiplet area or nominal bandwidth as sufficient proxies for manufacturing efficiency.

The reduced interconnect budget is matched by workload-aware parallelism. CASCADE adopts PDD to avoid layer-wise transfers between attention and MoE accelerators, and combines TP with either PP or context parallelism (CP) within the chiplet array. TP+PP localizes collectives to smaller chiplet groups and uses micro-batching to align compute demand with available memory bandwidth. TP+CP partitions KV caches across chiplets, avoiding KV cache replication and context-length-dependent communication for partial attention exchanges. These strategies trade off chiplet utilization, memory access, and D2D traffic differently as batch size and context length change. Across 69 feasible configurations in our parallelism DSE, neither TP-only nor PP-only achieves the best throughput, demonstrating why workload-specific execution is integral to a cost-constrained hardware design. Our main contributions are:
{\setlength{\leftmargini}{1.2em}
\begin{itemize}
    \setlength{\itemsep}{2pt}
    \item \textbf{Cost-aware, physically grounded chiplet-array design.}
    We introduce a homogeneous cHBM architecture with a workload-driven hardware design methodology that co-optimizes compute throughput, D2D bandwidth, HBM capacity, logic-die floorplanning, power, and manufacturing costs, making standard organic-substrate packaging viable under modeled constraints.

    \item \textbf{Workload-adaptive hardware--software co-design.} We combine PDD with TP+PP and TP+CP, selecting workload-tailored configurations through DSE to sustain high throughput and limit communication overhead under constrained D2D bandwidth.

    \item \textbf{Performance, energy, and manufacturing-cost evaluation.}
    CASCADE achieves up to $4.65\times$ higher decode throughput and $8.26\times$ higher energy efficiency than state-of-the-art accelerators. Parallelism DSE improves energy efficiency by up to $1.68\times$ over fixed execution on identical hardware. Our cost model estimates CASCADE's manufacturing cost per shipped package to be 38.4\% lower than that of AMMA under the evaluated assumptions.
\end{itemize}
}

\section{Background}\label{sec:2}
\subsection{Computation in MoE-LLM Decoding}\label{sec:2.1}
An MoE-LLM primarily comprises stacked Transformer decoder layers containing attention and MoE components. The attention layer includes query, key, and value (QKV) projections, attention computation, and output projection. During decoding, the arithmetic intensity of projection GEMMs increases with batch size through weight reuse. Attention variants differ: multi-head attention (MHA) uses general matrix-vector multiplications (GEMVs), GQA~\cite{gqa} uses thin GEMMs by sharing KV heads, and multi-head latent attention (MLA)~\cite{mla} uses GEMMs with compressed KV representations and relatively high arithmetic intensity. Unlike projections, decoding attention maintains approximately constant arithmetic intensity across batch sizes because sequences access independent KV caches.

The MoE layer comprises a router and FFNs. The router assigns tokens to expert FFNs~\cite{switchtransformers}, which perform gate, up, and down projections as grouped GEMMs. Because each expert processes only its routed tokens, its effective batch size and arithmetic intensity are lower than those of linear projections.

\subsection{Reticle Limits and Advanced Packaging}\label{sec:2.2}
\subsubsection{2.5D Packaging}\label{sec:2.2.1}
Semiconductor photolithography restricts die size to a single exposure field (reticle limit), making monolithic scaling difficult as high-performance computing demands grow. Conventional 2D packaging addresses this limit by integrating multiple dies on an organic package substrate, forming a multi-chip module (MCM)~\cite{mcm_gpu}. However, coarse substrate wiring pitch restricts interconnect density, while longer links increase signal delay and energy consumption. As chiplet counts rise, these constraints limit D2D bandwidth for communication-intensive workloads.

2.5D advanced packaging addresses these limitations using dense silicon interconnect structures. TSMC's chip-on-wafer-on-substrate (CoWoS)~\cite{cowos} can use a silicon interposer between chiplets and the substrate, enabling high-bandwidth communication over short distances but increasing packaging costs when large interposers are required. Alternatively, Intel's embedded multi-die interconnect bridge (EMIB)~\cite{emib} embeds small silicon bridges in the substrate, providing dense interconnections only where needed. Both approaches enable high-bandwidth D2D communication among chiplets.

\begin{figure}
\centering
 \includegraphics[width=1.0\linewidth]{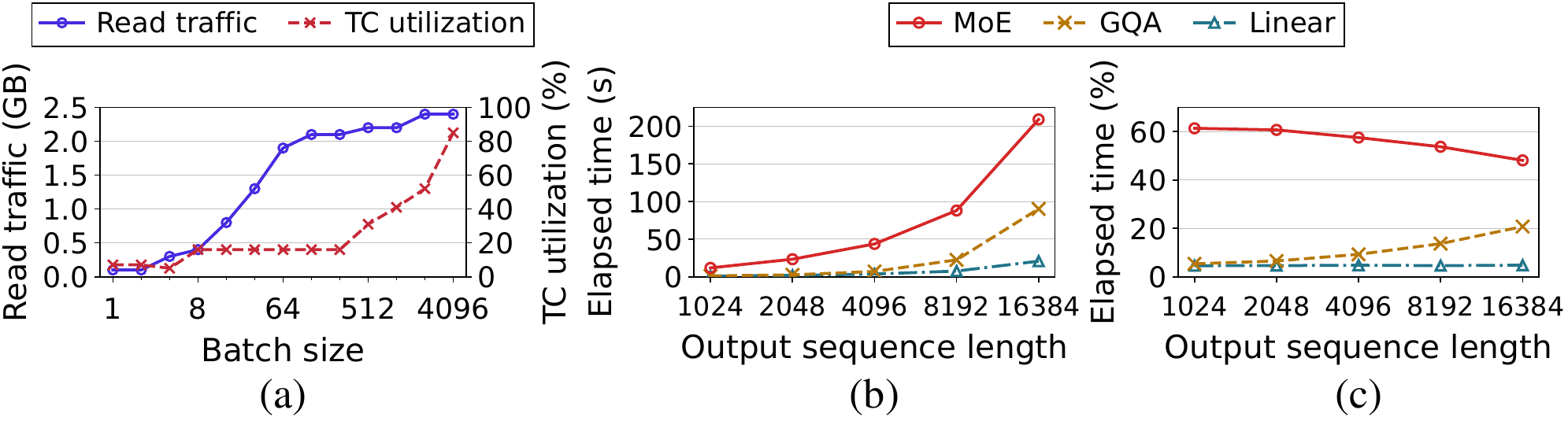}
\caption{Profiling results for GPT-OSS-120B. (a) DRAM read traffic and Tensor Core utilization of an MoE layer. (b) Wall-clock time and (c) percentage of total decode time for matmul kernels at batch size 32, excluding MoE router GEMMs.}
\label{fig:membound}
\end{figure}

\begin{figure}
\centering
 \includegraphics[width=1.0\linewidth]{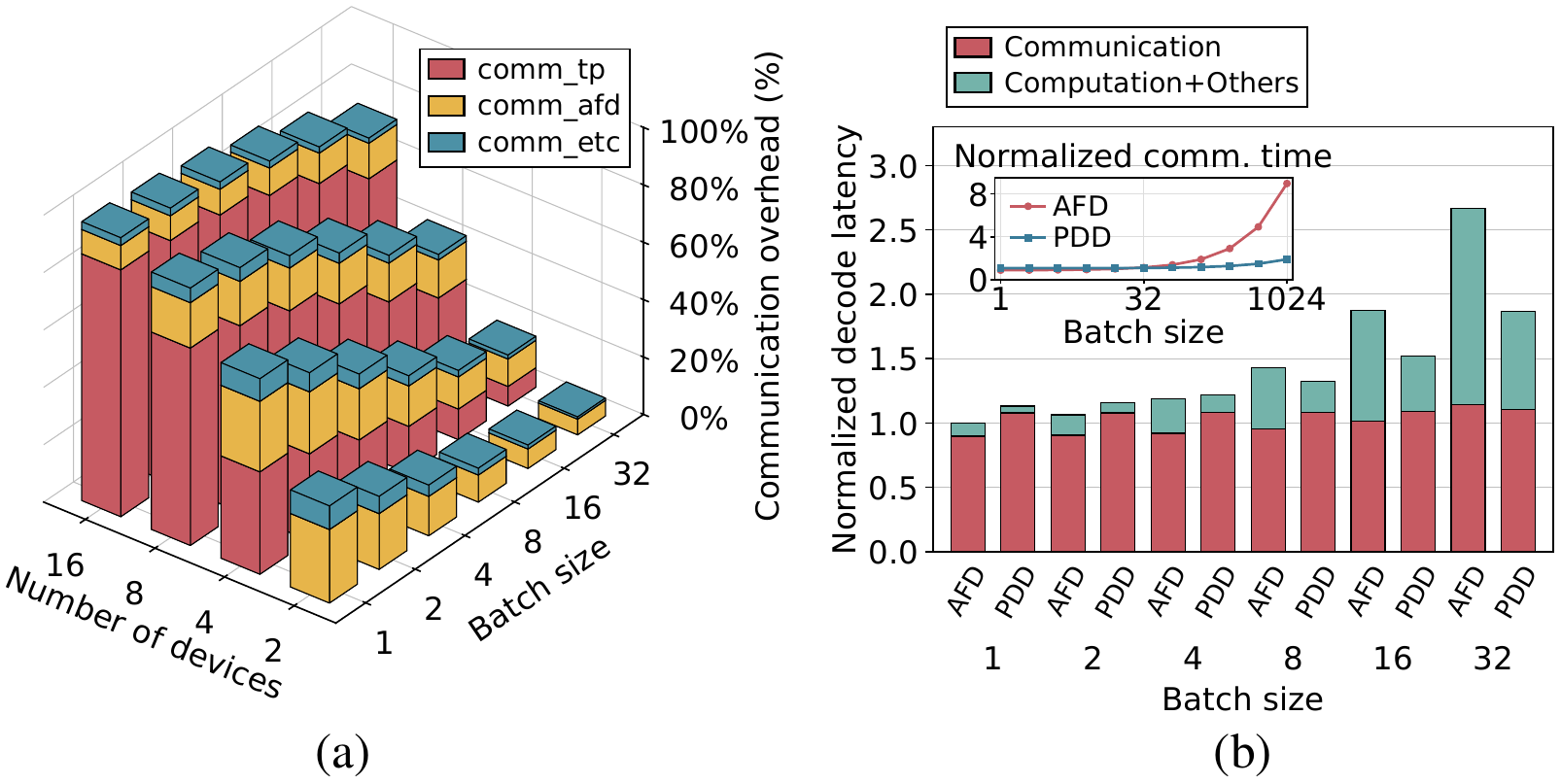}
\caption{(a) Communication overhead breakdown of AFD: \texttt{comm\_tp}, \texttt{comm\_afd}, and \texttt{comm\_etc} denote TP communication, activation transfers between attention and FFN groups, and other communication, respectively. (b) Decode latency breakdown of AFD and PDD, normalized to AFD at batch size 1.}
\label{fig:afd_comm}
\end{figure}

\subsubsection{3D Stacking}\label{sec:2.2.2}
Modern GPUs commonly use 2.5D packaging to integrate compute dies and HBM. However, the GPU and HBM remain separate components connected through package interconnects, and the GPU{\textendash}HBM interface limits usable bandwidth despite HBM's high internal bandwidth.

3D stacking addresses these limitations by vertically integrating compute and memory, with cHBM and 3D-stacked DRAM as representative approaches. Conventional HBM uses through-silicon vias (TSVs) to connect stacked core dies to an underlying logic die~\cite{hbm_tsv}. cHBM allows customized functions in this logic die~\cite{custom_hbm}. In compute-enabled cHBM, integrated processing units access DRAM directly through TSVs, exploiting HBM's internal bandwidth while reducing external data movement~\cite{hbm_pim,streamdq}. Alternatively, 3D-stacked DRAM integrates DRAM with a separate processing die, placing DRAM above or below logic~\cite{raptor_isca26,stratum,jssc_3ddram_2026,helios_micro26}. These architectures can use TSVs for vertical connections, while hybrid bonding provides dense connections between adjacent dies~\cite{hybrid_bonding}. Both approaches provide high memory bandwidth to processing units while reducing data transfer distances, benefiting memory-intensive AI workloads.

\section{Motivation}\label{sec:3}
\subsection{Memory-Bound Nature of LLM Decoding}\label{sec:3.1}
Although MoE-LLM decoding is often memory-bound in both attention and MoE computation, many prior memory-centric accelerators~\cite{hpu,amma} primarily target attention. Figure~\ref{fig:membound}(a) presents DRAM read traffic and Tensor Core utilization of Triton~\cite{triton} CUDA kernels for GPT-OSS-120B~\cite{gptoss} on an H100, profiled using Nsight Compute~\cite{nsight_compute}. Larger batches activate more experts, increasing DRAM reads with limited gains in arithmetic intensity. Tensor Core utilization remains below 16\% until the effective batch per expert grows sufficiently, confirming the MoE layer's memory-bound nature.

Figure~\ref{fig:membound}(b) and (c) present profiling results of major matrix multiplication (matmul) CUDA kernels in GPT-OSS-120B decoding with vLLM~\cite{vllm} on H100, measured using Nsight Systems~\cite{nsight_systems}. Although the arithmetic intensity of linear projections rises with batch size through weight reuse, they account for less than 5\% of decode time at batch 32 across evaluated output lengths. Memory-bound GQA and MoE grouped GEMMs dominate decode time, motivating acceleration of both attention and MoE.

\subsection{Rationale for Prefill{\textendash}Decode Disaggregation}\label{sec:3.2}
Figure~\ref{fig:afd_comm} compares AFD communication overhead and AFD/PDD decode latency for GPT-OSS-120B at a 100K final sequence length using analytical modeling. We assume xPUs with Rubin-equivalent compute throughput and memory bandwidth~\cite{rubin}, NVLink-based chip-to-chip (C2C) links within groups, and host-mediated PCIe between AFD's equally sized attention and FFN groups. Figure~\ref{fig:afd_comm}(a) shows that AFD adds activation-transfer overhead beyond TP communication, while total communication increasingly dominates latency as device count grows, especially at smaller batch sizes. Figure~\ref{fig:afd_comm}(b) compares single-batch latency on eight total devices, rather than continuous batch streaming. Because AFD divides devices between groups, fewer devices handle each operation than in PDD, increasing computation latency. Although AFD is slightly faster at very small batches, PDD becomes increasingly advantageous as batch size grows. These results motivate adopting PDD for CASCADE's latency-sensitive decoding across batch sizes 1--32.

\subsection{GPU Shoreline Limitations}\label{sec:3.3}
Memory bandwidth can scale by increasing bandwidth per HBM stack (scale-up) or the number of stacks (scale-out). While scale-up depends on HBM advances, scale-out can exploit existing stacks but is constrained by conventional 2.5D GPU{\textendash}HBM packaging.

These packages integrate GPUs and HBM stacks on a silicon interposer~\cite{cowos,hbm_interposer}, with GPU die size constrained by the reticle limit~\cite{mcm_gpu}. Each HBM stack requires a wide, short-reach GPU interface, consuming HBM PHY area and limited GPU shoreline~\cite{memory_fabric}. Adding stacks increases demand for shoreline and interconnect routing resources, but expanding the interposer does not increase GPU shoreline. Consequently, aggregate memory bandwidth scaling through scale-out remains constrained.

\begin{figure*}
\centering
 \includegraphics[width=1.0\linewidth]{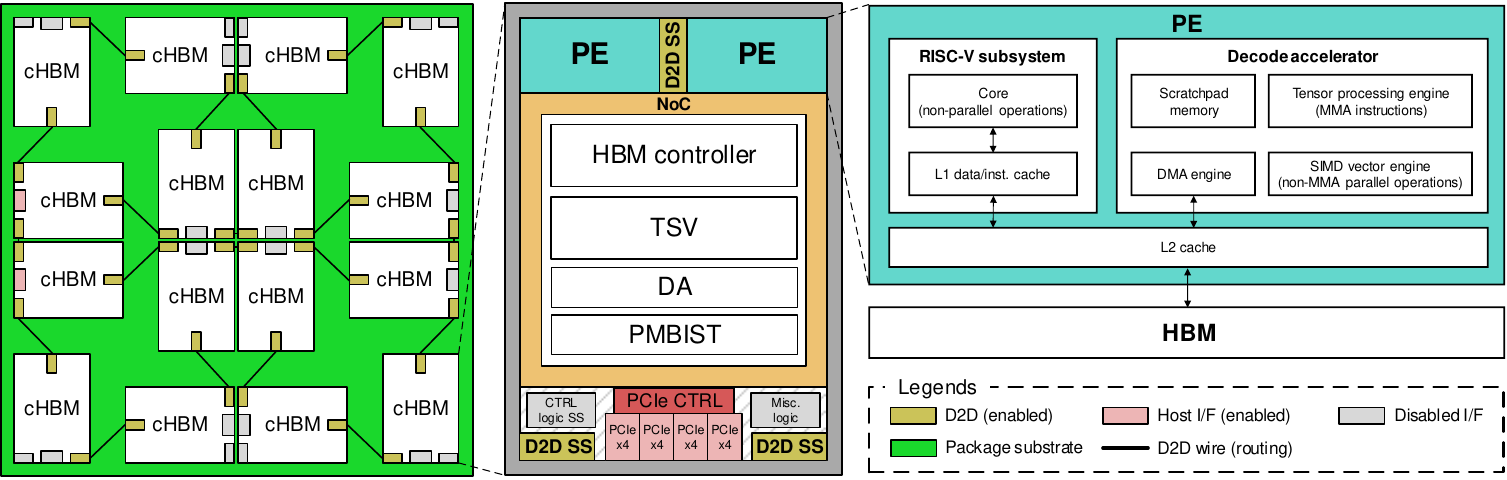}
\caption{CASCADE architecture overview. Top view of the CASCADE device composed of cHBM chiplets; logic die subsystem within each cHBM; and PE subsystem within the logic die. The geometric representation of the components reflects a schematic layout with placement informed by physical layout considerations, while the depicted component sizes do not exactly reflect their actual proportions and only the key architectural blocks are shown.}
\label{fig:cascade_topview}
\end{figure*}

\subsection{Breakthrough via 3D Stacking}\label{sec:3.4}
3D stacking enables PNM accelerators to exploit HBM's high internal bandwidth via vertical links, bringing computation closer to data. Duplex~\cite{duplex} combines an xPU with Logic-PIM for attention and MoE, while Stratum~\cite{stratum} uses hybrid-bonded 3D-stacked DRAM for MoE serving. Other hybrid-bonding-based accelerators~\cite{hb_npu,h2llm,compair_isca26} leverage dense DRAM-to-logic connections to accelerate matmul.

At the package level, compute-enabled cHBM can overcome GPU shoreline limitations by replacing the GPU die with interconnected cHBM chiplets. Each chiplet contains processing units that access its local DRAM through vertical connections, eliminating the need for additional HBM PHYs on a central GPU. More HBM stacks can be integrated within a comparable package footprint, allowing aggregate memory bandwidth to scale with stack count rather than GPU shoreline. AMMA~\cite{amma} demonstrates this approach using HBM-PNM chiplets to accelerate memory-bound attention.

\subsection{Limitations of State-of-the-Art Accelerators}\label{sec:3.5}
Duplex~\cite{duplex} retains GPU shoreline constraints, limiting aggregate memory bandwidth scaling by adding HBM stacks. To accelerate memory-bound attention and MoE on Logic-PIM, Duplex instead expands per-stack internal bandwidth fourfold through TSVs, introducing routing complexity and area and power overheads~\cite{duopd}. Extending this design from HBM3 to HBM4 requires higher compute throughput to maintain the compute-to-bandwidth ratio, increasing logic-die area and power. Otherwise, memory-bound operations may become compute-bound on Logic-PIM, limiting acceleration.

AMMA~\cite{amma} employs 16 chiplets with high compute throughput and four-directional D2D interfaces, but its hardware specifications are not sufficiently justified in terms of workload requirements, area, power, and packaging costs. In particular, limited performance gains from increasing D2D bandwidth suggest that its provisioned bandwidth may be excessive relative to PHY area and packaging costs. Moreover, HBM logic dies must accommodate TSVs, memory PHYs, design-for-test (DFT) logic, and direct-access (DA) test interfaces~\cite{samsung_hbm4}, limiting available area for compute units and D2D interfaces. The physical feasibility and manufacturing costs of integrating 16 chiplets remain insufficiently addressed.

AMMA employs AFD with fixed parallelism, offloading attention to its chiplets and FFN or MoE computation to separate GPUs or LPUs. Alternative execution strategies could improve hardware utilization across workloads, but without exploring them, it remains unclear whether AMMA's hardware specifications are well matched to workload requirements. These considerations motivate CASCADE to jointly consider manufacturing cost, physical feasibility, and workload-aware execution strategies.

\subsection{Cost-Aware Accelerator Design}\label{sec:3.6}
This work proposes a new architecture that prioritizes manufacturing cost reduction and physical feasibility while maximizing performance through software and hardware co-design. Increasing HBM stack height requires more DRAM core dies and greater stacking complexity, raising manufacturing costs and yield risks~\cite{hbm_packaging,hbm_tsv_yield,hotchips_hbm}. For HBM4, a 4-Hi stack provides full nominal per-stack bandwidth, while additional core dies increase capacity without increasing bandwidth~\cite{semianalysis_4hi}. When capacity accommodates model weights and KV caches, reducing HBM stack height can lower manufacturing costs and power consumption~\cite{hbm_power} without sacrificing bandwidth.

At the package level, high-bandwidth D2D interfaces may require costly advanced packaging, such as silicon interposers~\cite{chipletcost}. However, provisioning bandwidth beyond workload needs can increase implementation costs without proportional performance gains. For workloads less sensitive to D2D bandwidth, reducing it can enable cost-effective standard organic substrate packaging if interconnect density and signal integrity requirements are satisfied.

\section{CASCADE Architecture}\label{sec:4}
\subsection{Overall Architecture}\label{sec:4.1}
To overcome the memory bandwidth limitation of conventional GPU{\textendash}HBM architectures, we propose CASCADE, a memory-centric accelerator composed entirely of cHBM chiplets. Figure~\ref{fig:cascade_topview} illustrates the architecture. Instead of a single large compute die, CASCADE scales memory bandwidth through an array of cHBM chiplets, each integrating processing elements (PEs) within the cHBM logic die.

CASCADE comprises 16 cHBM chiplets in a 4 $\times$ 4 array. Each chiplet provides UCIe 3.0 standard package (UCIe-S)~\cite{ucie_3_0} D2D interfaces, each with 128 GB/s per direction, and a host interface of PCIe 6.0 x16 (64 GT/s per lane, 256 GB/s bidirectional bandwidth)~\cite{pcie_6_0}. Two chiplets serve as master chiplets, communicating directly with the host, while other chiplets communicate only through D2D links.

Each cHBM uses an HBM4 8-Hi stack, providing 24 GB capacity, 2.75 TB/s memory bandwidth, and peak throughputs of 22 TFLOPS for BF16 tensor operations and 256 GFLOPS for FP32 vector operations. In addition to the HBM controller and control logic, the logic die integrates two PEs dedicated to LLM decoding. These components communicate through a network-on-chip (NoC). Each PE accesses data from its local HBM stack through TSVs, while data exchanges between chiplets use D2D links. By integrating computation within the logic die, CASCADE minimizes data movement while exploiting the chiplet array's aggregate memory bandwidth.

Each PE consists of a RISC-V~\cite{riscv_isa} subsystem and a decode accelerator. The RISC-V subsystem handles control flow and non-parallel operations. The decode accelerator includes a tensor processing engine for matrix multiply-accumulate (MMA) operations, a SIMD vector engine for non-MMA operations including softmax-related computations, and a 2 MiB SRAM scratchpad. The scratchpad provides software-managed storage for data reuse during MMA.

\subsection{Homogeneous Chiplet Design Rationale}\label{sec:4.2}
The physical layout of the cHBM logic die constrains the placement of D2D interfaces. As shown in Figure~\ref{fig:cascade_topview}, D2D interfaces can be accommodated along the east, west, and north sides through area optimization. However, the south side accommodates the host interface, its controller, and miscellaneous logic, leaving insufficient area for an additional D2D interface. Consequently, each chiplet can support D2D communication in at most three directions. AMMA~\cite{amma} similarly arranges 16 HBM-PNM chiplets in a $4\times4$ array but integrates four-directional D2D interfaces within each logic die. However, it does not establish how these interfaces can coexist with host communication circuitry within the limited logic die area. Therefore, this four-directional configuration is impractical under the physical constraints of our cHBM floorplan.

Supporting four-directional D2D communication requires alternative chiplet configurations. One option is to dedicate host interfaces only to the master chiplets, allowing the remaining chiplets to support four-directional D2D interfaces. Another option is to remove the host interfaces from all cHBM chiplets, use a separate I/O die for host communication, and provide four-directional D2D interfaces for all chiplets. Thus, AMMA would require a separate I/O die to maintain four-directional D2D interfaces on all chiplets. However, such heterogeneous chiplet configurations increase manufacturing cost and packaging complexity~\cite{chipletcost}.

Therefore, CASCADE adopts a homogeneous chiplet design. Although only the master chiplets utilize the host interface, all cHBM chiplets share an identical physical structure. CASCADE accommodates the resulting three-directional D2D constraint through the efficient parallel execution strategies described in Section~\ref{sec:5}.

\subsection{Hardware Specification Rationale}\label{sec:4.3}
\subsubsection{Compute Throughput}\label{sec:4.3.1}
\begin{figure}
\centering
 \includegraphics[width=1.0\linewidth]{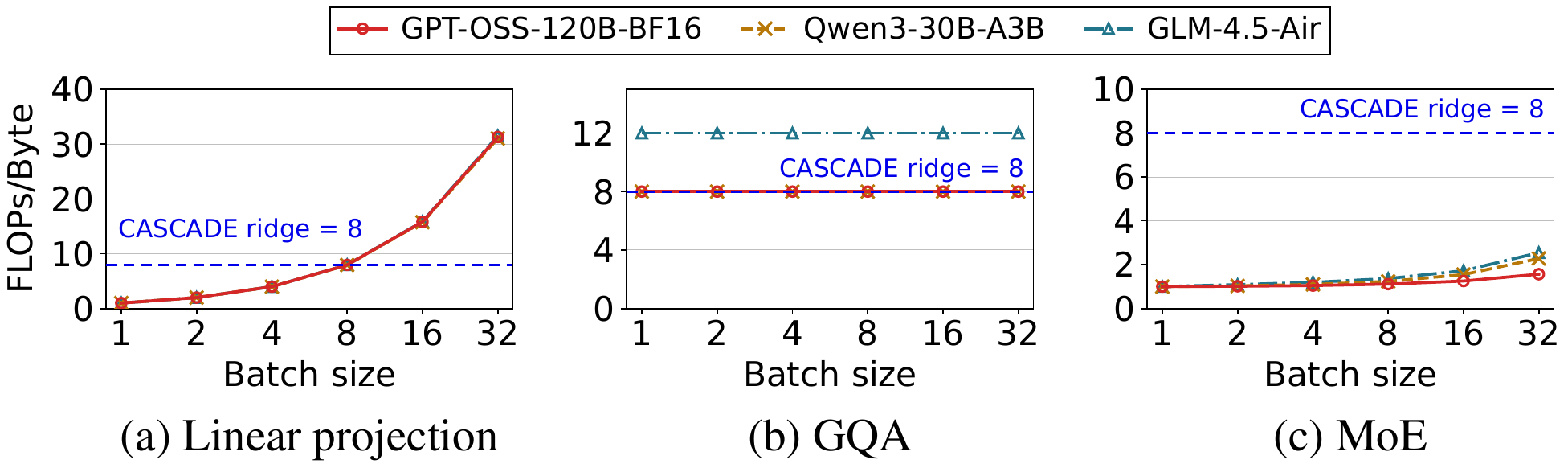}
\caption{Arithmetic intensity of (a) linear projection, (b) GQA, and (c) MoE during decoding across batch sizes and models. The horizontal dashed line indicates the CASCADE ridge point of 8 FLOPs/Byte.}
\label{fig:arithmetic_intensity}
\end{figure}

We determine CASCADE's tensor processing engine throughput from the arithmetic intensity of major matmul operations: linear projections, GQA, and MoE grouped GEMMs. Figure~\ref{fig:arithmetic_intensity} compares workloads across three models~\cite{gptoss,qwen3,glm45}. All operations use BF16 computation, with weights, activations, and KV caches stored in BF16. The arithmetic intensity of linear projections increases approximately linearly with batch size, approaching 32 FLOPs/Byte at batch size 32. In contrast, GQA's arithmetic intensity approximately equals the Q-to-KV head ratio: 8 for GPT-OSS-120B-BF16 and Qwen3-30B-A3B, and 12 for GLM-4.5-Air. These values remain approximately constant.

MoE grouped GEMMs instead exhibit routing-dependent arithmetic intensity. We therefore use a probabilistic model to account for the distinct experts activated within a batch. Let $b$ denote the batch size, $e$ the total number of experts available for routing, and $k$ the number of experts selected per token. Assuming uniform expert selection and independent routing across tokens, the expected number of activated experts is $\bar{e}_{\mathrm{act}} = e[1-(1-k/e)^b]$. We estimate MoE grouped GEMM arithmetic intensity using the expected number of activated experts, following prior arithmetic intensity analyses~\cite{sonicmoe,nvmatmul}. For equally sized routed and shared experts, let $n_s$ denote the number of shared experts, $h$ the hidden dimension, $d$ the expert intermediate dimension, and $s_w=2$ the storage bytes per BF16 weight. With $t=b(k+n_s)$ token-expert assignments, we estimate the arithmetic intensity assuming one-time transfers of each GEMM's weights, inputs, and outputs:
\begin{equation}
I_{\mathrm{MoE}}
\approx
\frac{2thd}
{s_w hd(\bar{e}_{\mathrm{act}}+n_s)+2t(h+d)}.
\label{eq:moe_ai}
\end{equation}
As shown in Figure~\ref{fig:arithmetic_intensity}(c), modeled grouped GEMM arithmetic intensity is below 8 FLOPs/Byte across batch sizes for all three models.

We address the higher arithmetic intensity of linear projections via micro-batching (Section~\ref{sec:5}). Splitting a batch of 32 into four micro-batches lowers their arithmetic intensity to below 8 FLOPs/Byte for all three models. Accordingly, we provision 22 TFLOPS of peak BF16 compute throughput per cHBM chiplet. Combined with 2.75 TB/s HBM bandwidth, this yields a ridge point of 8 FLOPs/Byte. Under micro-batching, the analyzed workloads operate at or below this ridge point, except for GQA in GLM-4.5-Air.

PE scratchpad capacity also affects attention efficiency, as insufficient capacity limits KV block size~\cite{flashattention}. We use 128-token blocks to reduce online softmax overhead, allowing attention to approach the HBM bandwidth bound. Therefore, we provision 2 MiB SRAM per PE for these blocks and intermediate data. This capacity satisfies area and power budgets estimated from a state-of-the-art NPU~\cite{rebel_isscc26}.

\subsubsection{D2D Bandwidth}\label{sec:4.3.2}
CASCADE considers standard packaging to reduce packaging cost. Standard packaging uses a coarser bump pitch and provides lower D2D bandwidth density than advanced packaging~\cite{ucie_packaging}. Consequently, achieving the same bandwidth requires a larger D2D interface footprint. Within our cHBM logic die floorplan, accommodating all required D2D lanes under these constraints limits the maximum D2D bandwidth to 128 GB/s.

\begin{figure}
\centering
 \includegraphics[width=1.0\linewidth]{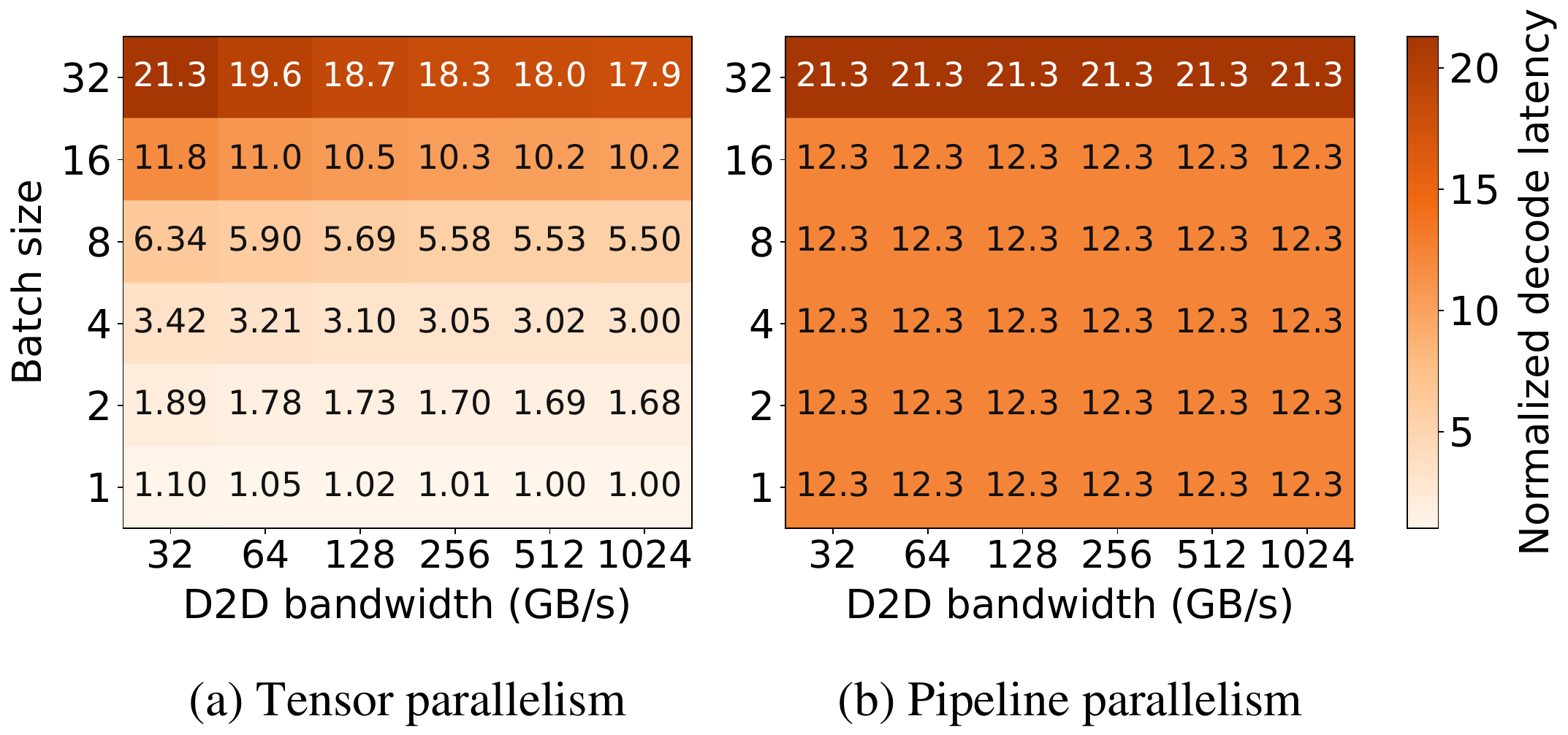}
\caption{D2D bandwidth sensitivity for GPT-OSS-120B-BF16. Latency is normalized to tensor parallelism at batch size 1 and 1024 GB/s D2D bandwidth. Memory capacity constraints are not enforced in this bandwidth sensitivity analysis.}
\label{fig:d2d_dse}
\end{figure}

Using analytical modeling, we explore the design space of D2D bandwidth and batch size for GPT-OSS-120B-BF16. Figure~\ref{fig:d2d_dse} presents the decode latency under tensor parallelism (TP) and pipeline parallelism (PP) with a final sequence length of 100K tokens. We focus on TP because it requires frequent collective communication between chiplets~\cite{megatronlm}. In contrast, PP primarily transfers activations at pipeline stage boundaries and exhibits limited sensitivity to D2D bandwidth. TP achieves lower decode latency than PP in all explored configurations except at batch size 32 and 32 GB/s D2D bandwidth. By enabling all chiplets to run in parallel at any batch size, TP improves chiplet utilization, motivating us to retain TP as one dimension of our two-dimensional (2D) parallelism strategies.

Under TP, lower D2D bandwidth increases communication latency for activation transfers between chiplets. Across batch sizes 1--32, reducing bandwidth from 1024 GB/s to 128 GB/s increases decode latency by approximately 2.4--4.2\%, whereas reducing it to 64 GB/s increases latency by approximately 5.1--9.0\%. Therefore, we select 128 GB/s to limit the degradation in small-batch performance while satisfying the physical constraints of standard packaging.

\subsubsection{Memory Capacity}\label{sec:4.3.3}
Quantized weights can reduce the memory footprint of models that exceed the capacity of 4-Hi stacks. However, CASCADE's BF16 tensor processing engines require weight dequantization for GEMM execution. Although fused kernels can overlap HBM reads, SIMD dequantization, and tensor processing engine execution across tiles~\cite{marlin,liquidgemm}, limited SIMD vector engine throughput can become a bottleneck. Substantially increasing SIMD throughput to match the weight supply rate would strain the logic die's area and power budgets. Therefore, CASCADE stores weights in BF16, matching its compute precision and eliminating runtime weight dequantization without increasing SIMD throughput.

Storing weights in BF16 increases memory capacity requirements, making 4-Hi stacks insufficient for the larger target models. Stacks with more core dies provide additional capacity for larger models and longer sequences, but increase manufacturing costs and memory power consumption~\cite{hbm_packaging,hbm_power}. We therefore adopt HBM4 8-Hi stacks to balance these capacity benefits against cost and power overheads, providing 24 GB per chiplet and 384 GB per device.

\begin{table}[t]
\caption{Area budget per chiplet for CASCADE and AMMA.}
\label{tab:area_estimation}
\centering
\footnotesize
\setlength{\tabcolsep}{2pt}
\renewcommand{\arraystretch}{1.04}
\newcommand{\chipletSectionRule}{\noalign{\hrule height 0.9pt\vskip 2pt}}
\begin{tabular*}{\columnwidth}{@{\extracolsep{\fill}}lrrr@{}}
\chipletSectionRule
\multicolumn{4}{@{}l}{\textbf{Reserved area for HBM IPs and Keep-Out Zone}}\\
Block & $W$ (mm) & $H$ (mm) & Area (mm$^2$)\\
\hline
HBM data TSVs~\cite{samsung_hbm4}           & 10 & 2.2 & 22\\
HBM DA                  & 10 & 1.8 & 18\\
HBM MBIST \& IEEE 1500  & 10 & 2.4 & 24\\
HBM controller and PHY~\cite{samsung_chbm4}          & 10 & 2.5 & 25\\
P/G TSV KOZ~\cite{pgtsv_koz} & 10 & 11.8 & 3.54\\
Die-boundary KOZ (depth: 0.5mm)       & $W$ & $H$ & $2\times0.5\times(W+H)$\\
\multicolumn{3}{@{}l}{Total reserved area} & $92.54+W+H$\\
\hline
\end{tabular*}
\par\medskip
\begin{tabular*}{\columnwidth}{@{\extracolsep{\fill}}lrrr@{}}
\chipletSectionRule
\multicolumn{4}{@{}l}{\textbf{Common IP dimensions and unit areas}}\\
IP & $W$ (mm) & $H$ (mm) & Area (mm$^2$)\\
\hline
$2\times$ UCIe-S ($\times16$), stacked, 32Gbps~\cite{ucie_3_0, ucie_packaging, uciephy_eliyan, uciephy_isscc1, uciephy_isscc2} & 1.14 & 1.64 & 1.8696\\
$2\times$ UCIe-S ($\times16$), unstacked, 32Gbps & 2.28 & 0.82 & 1.8696\\
UCIe-A ($\times64$), 32Gbps            & 0.4  & 1.45 & 0.58\\
UCIe controller              & -    & -    & 0.5\\
PCIe ($\times4$), 64 GT/s/lane~\cite{pcie_6_0, pciephy_pcie5, pciephy_pam4_1, pciephy_pam4_2} & 1.35 & 1.35 & 1.8225\\
PCIe controller per 4 lanes  & -    & -    & 1\\
Control \& Misc. IP  & - & - & 10\\
\hline
\end{tabular*}
\par\medskip
\begin{tabular*}{\columnwidth}{@{\extracolsep{\fill}}lrrr@{}}
\chipletSectionRule
\multicolumn{2}{@{}l}{\textbf{Area by chiplet (mm$^2$)}} &
\multicolumn{2}{c@{}}{AMMA~\cite{amma}}\\
\cline{3-4}
IP & CASCADE & Compute & I/O\\
\hline
$2\times$ UCIe-S ($\times16$), unstacked & 5.6088  & -     & -    \\
UCIe-A ($\times64$)            & -       & 13.92 & 3.48\\
UCIe controller              & 1.5     & 12    & 3 \\
PCIe ($\times4$), 64 GT/s per lane & 7.29 & -   & 7.29\\
PCIe controller per 4 lanes  & 4       & -     & 4\\
PE + SRAM                    & 18.3    & 17.93~\cite{tsmc2021n4p,tsmcN3eDensity} & - \\
NoC                          & 25      & 25    & -    \\
Control \& Misc. IP          & 10      & 10    & 10    \\
Side routing~\cite{routing_floonoc,routing_intel} & 2.2784 & 26.70 & 0\\
\hline
Required IP and routing area & 73.9772  & 105.55 & 27.77\\
Boundary KOZ                 & 28.0     & 30.4   & 12.8\\
\textbf{Total required area} & \textbf{194.5172} & \textbf{228.49} & \textbf{40.57}\\
Selected $H\times W$ (mm)    & $15.2\times12.8$ & $15.4\times15$ & $6\times6.8$\\
Die area ($H W$)             & 194.56 & 231.00 & 40.80\\
\hline
\end{tabular*}
\par\smallskip
\begin{tabular*}{\columnwidth}{@{\extracolsep{\fill}}lrl@{}}
\chipletSectionRule
\multicolumn{3}{@{}l}{\textbf{Key manufacturing cost inputs (USD)}}\\
Cost input & Value & Unit\\
\hline
3-nm logic/I/O wafer~\cite{epoch_3nm_wafer_cost} & 20,000 & /wafer\\
Standard substrate~\cite{accelstack} & 0.005 & /mm$^2$\\
CoWoS-R~\cite{goldstate2025packaging} & 0.074 & /mm$^2$\\
CoWoS-L~\cite{ifast2026packaging} & 0.390 & /mm$^2$\\
EMIB-T~\cite{ifast2026packaging} & 0.250 & /mm$^2$\\
\hline
\end{tabular*}
\par\smallskip
\begin{minipage}{\columnwidth}
\scriptsize
We assume 3\% P/G TSV coverage and a 0.5-mm die-boundary KOZ. We scale AMMA's 4-nm MAC area to $11.82 (1.06/1.60)\approx7.83$~mm$^2$ and budget 10.1~mm$^2$ for SRAM (10.01~mm$^2$ reported).
\end{minipage}
\end{table}

\subsection{Cost-Aware Die and Package Co-Design}\label{sec:4.4}
We conduct a preliminary exploration to identify an optimal point for a more practical SiP-level architecture, considering physical feasibility and fabrication-level cost efficiency. We then select the most cost-efficient design and compare it with AMMA\cite{amma}.

We modify the cost model based on AccelStack~\cite{accelstack}, using its yield, test, and non-recurring engineering (NRE) coefficients. To calculate recurring engineering (RE) costs more realistically, we adopt market estimates of 3-nm wafer and advanced packaging costs~\cite{epoch_3nm_wafer_cost, goldstate2025packaging,ifast2026packaging} (Table~\ref{tab:area_estimation}). Using \$20{,}000 as the baseline 3-nm wafer price, we model a 2.9\% price increase for every 10\% increase in exposures per wafer relative to the reference layout~\cite{semian_die_recticle, imec_mfu}.

\subsubsection{Die Size}
We select the logic-die dimensions that minimize the cost per known-good die (KGD) among floorplans subject to physical constraints imposed by packaging.

\begin{figure}
  \centering
  \includegraphics[width=\linewidth]{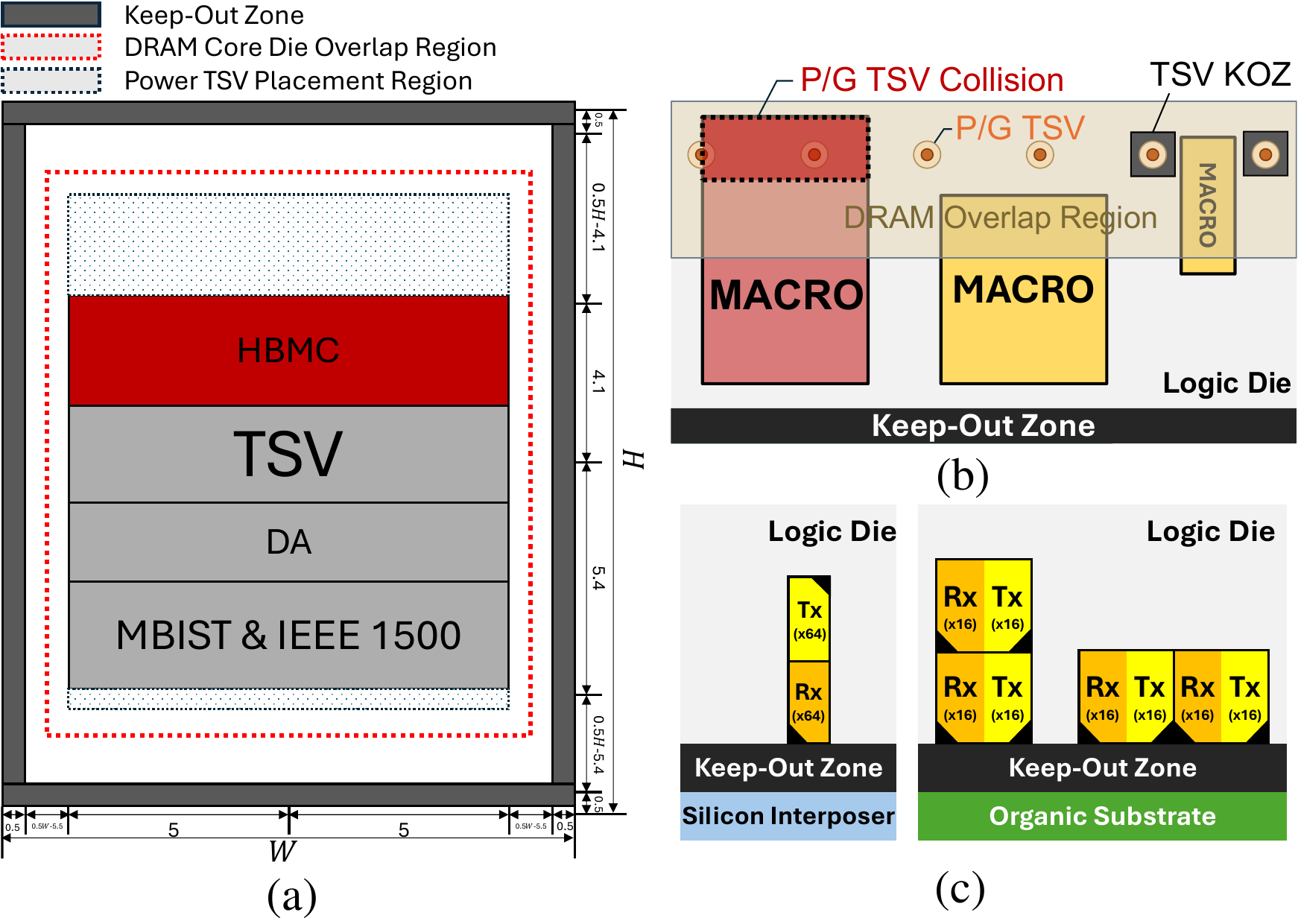}
  \caption{Die size and floorplan exploration: (a) logic die before
    floorplanning, (b) macro placement constraints from P/G TSVs,
    and (c) UCIe-A vs.\ UCIe-S (stacked/unstacked).}
  \label{fig:floorplan_dse}
\end{figure}

\noindent\textbf{Floorplan Feasibility.}
The standard HBM logic die requires IPs to support its memory functions and RAS features, as shown in Figure~\ref{fig:floorplan_dse}(a). Based on HBM4 die shots~\cite{samsung_hbm4, samsung_chbm4}, we estimate the reserved area for these IPs and the Keep-Out Zones (KOZs) at the chiplet boundary (Table~\ref{tab:area_estimation}). To enable a practical HBM-stacked 3D-IC, the total required area should include the reserved area of $(92.54+W+H)$~mm$^2$. We also consider the physical constraints of 3D packaging: Power/Ground (P/G) TSVs hinder the placement of large macros such as PHYs under the DRAM overlap region (Figure~\ref{fig:floorplan_dse}(b)). Therefore, we search for floorplans that avoid placing PHYs in the DRAM core die overlap region.

To select the optimal die dimension, we assume CASCADE and AMMA-Compute use the same area for NoC, control, and miscellaneous IPs, based on a rough estimate using 3-nm commercial IPs. CASCADE and AMMA-Compute have nearly identical combined PE and SRAM areas, so the largest area difference arises from their D2D bandwidth requirements. The required D2D bandwidth affects routing area and packaging selection, which depends on shoreline length and UCIe interface area. AMMA adopts 1500 GB/s per edge, necessitating advanced packaging and UCIe-A. However, CASCADE requires only 128 GB/s, enabling standard packaging for SiP. To satisfy the height constraint between the KOZ and the upper-side P/G TSV region, we select an unstacked 2$\times$ UCIe-S ($\times$16) module configuration (Figure~\ref{fig:floorplan_dse}(c)) supported by the UCIe specification~\cite{ucie_3_0}. As a result, CASCADE requires a 194.56 mm$^2$ die, whereas AMMA-Compute requires 231.00 mm$^2$.

\begin{figure}
\centering
 \includegraphics[width=1.0\linewidth]{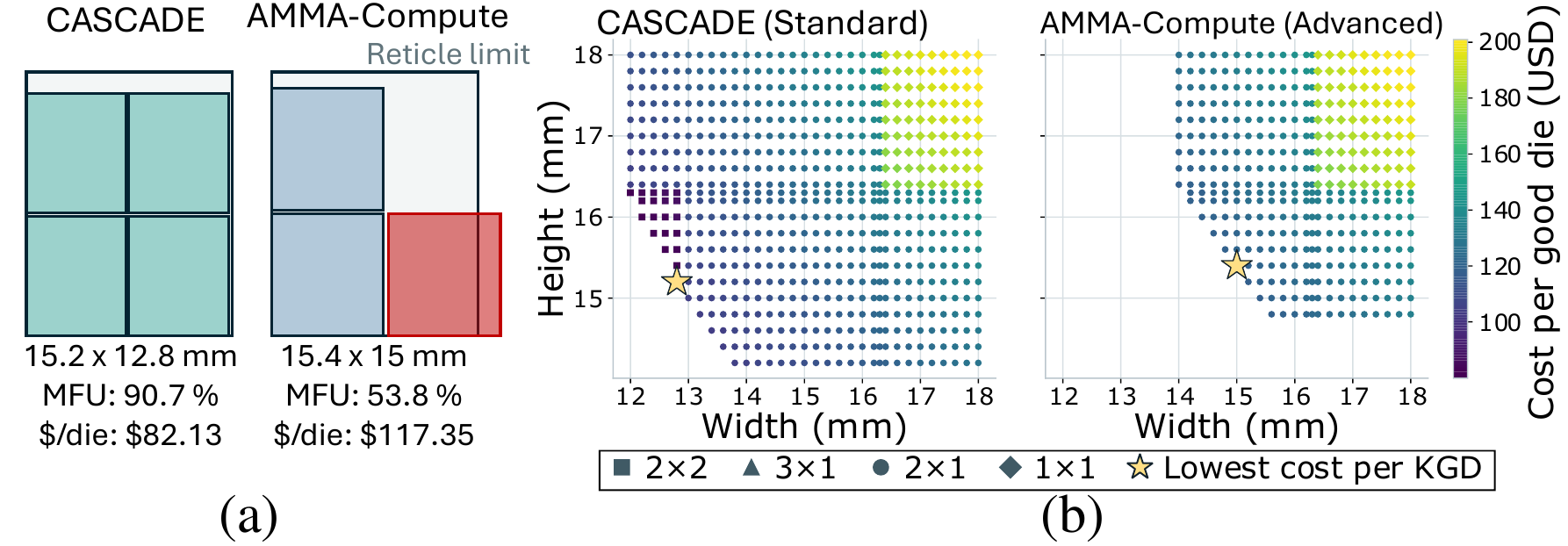}
\caption{Cost-efficient die size exploration: (a) relationship between die size and MFU, and (b) relative price per die.}
\label{fig:die_dse}
\end{figure}

\noindent\textbf{Fabrication Cost.}
The price per die depends primarily on the number of dies per wafer, but a suboptimal die dimension also increases manufacturing cost. As shown in Figure~\ref{fig:die_dse}(a), the $26 \times 33$~mm reticle field cannot accommodate four AMMA-Compute chiplet dies at the optimal dimensions. This decreases the Mask Field Utilization (MFU). Lower MFU increases the number of shots in the photo scanner process, leading to higher manufacturing costs~\cite{semian_die_recticle,imec_mfu,jeong2011mask}. Therefore, we calculate the price per die using the exposure-adjusted wafer cost to find the best dimension for CASCADE. Figure~\ref{fig:die_dse}(b) shows why maximizing the die area within a reticle field is insufficient. As a result, we select the most cost-efficient die dimensions of $15.2 \times 12.8$~mm, which yield a 90.7\% MFU and a cost of \$82.13 per die (Figure~\ref{fig:die_dse}).

\begin{figure}
\centering
 \includegraphics[width=1.0\linewidth]{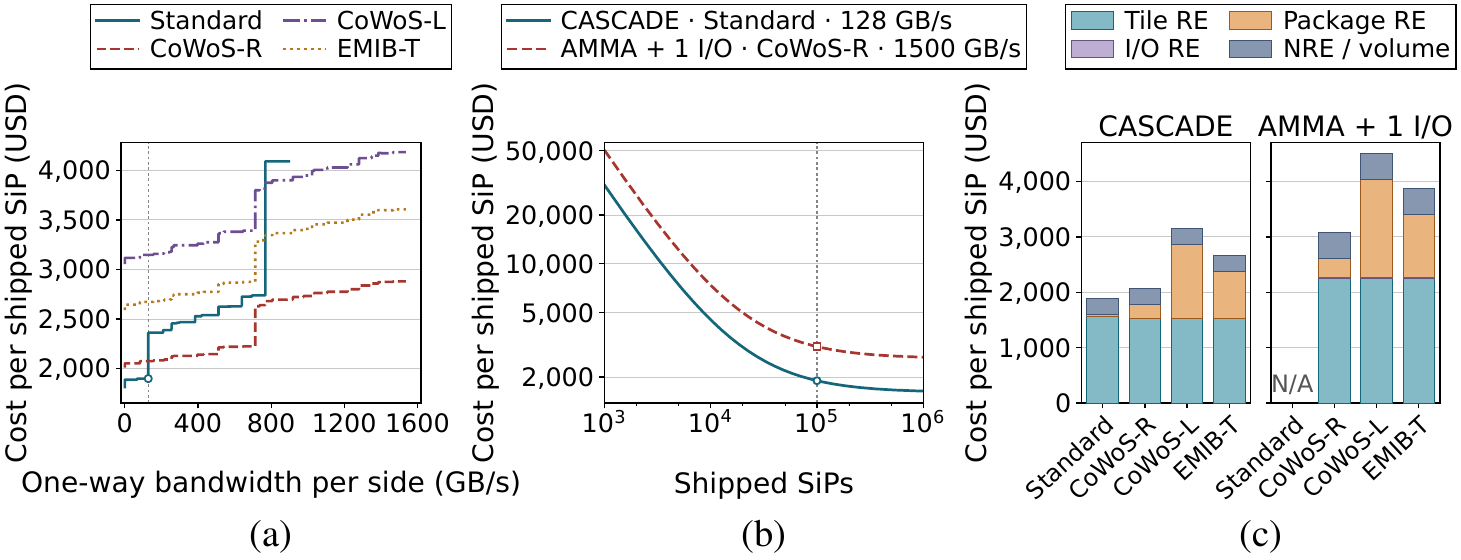}
\caption{Manufacturing cost per shipped SiP: (a) CASCADE cost vs.\ link bandwidth, (b) CASCADE and AMMA costs vs.\ volume, and (c) cost breakdown by packaging technology.}
\label{fig:sip_cost}
\end{figure}

\begin{figure}
\centering
 \includegraphics[width=1.0\linewidth]{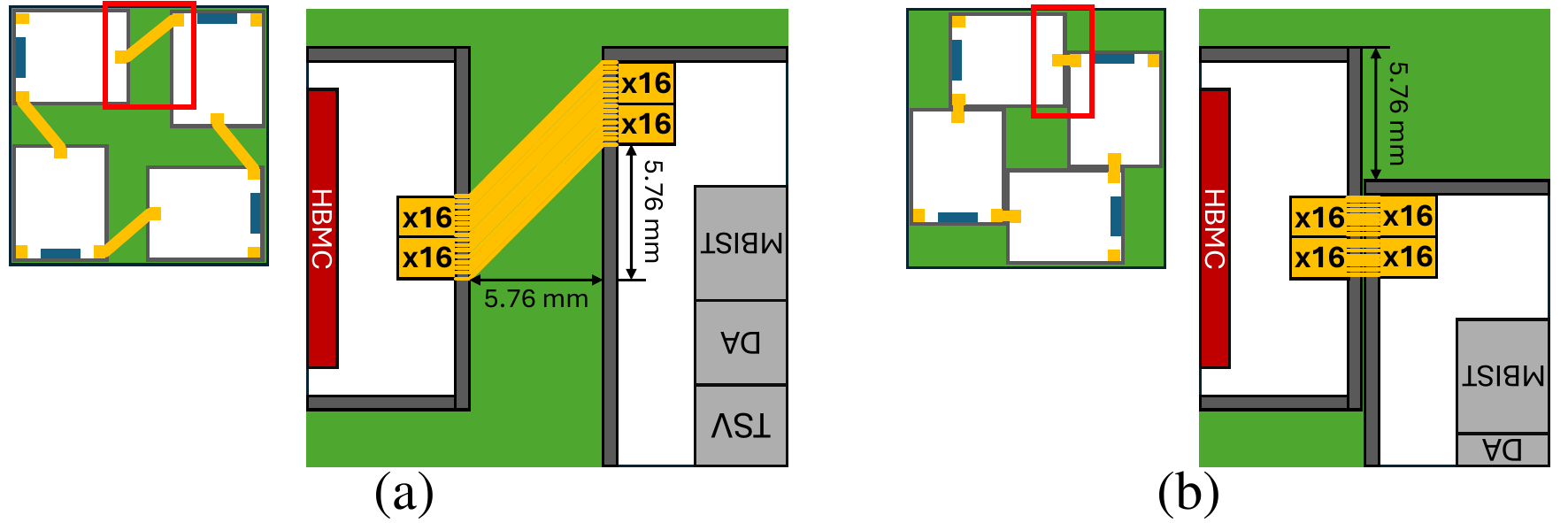}
\caption{Die placement and package routing: (a) aligned dies with diagonal links and (b) staggered dies with facing PHYs.}
\label{fig:sip_dse}
\end{figure}

\subsubsection{Package Selection and Die Placement}
Standard packaging offers low manufacturing costs but requires substantially more die area for the same D2D bandwidth. In contrast, advanced packaging guarantees high D2D bandwidth with small area consumption but incurs high manufacturing costs. To identify a cost-effective bandwidth range and justify the choice of standard packaging for CASCADE, we sweep the per-edge D2D bandwidth and calculate the SiP cost. In Figure~\ref{fig:sip_cost}(a), we compare standard packaging with CoWoS-R, CoWoS-L~\cite{tsmcCowos}, and EMIB-T~\cite{intel2025emib}. At 100{,}000 shipped SiPs, standard packaging is the lowest-cost option for CASCADE at D2D bandwidths up to and including 128~GB/s per edge. However, above 128~GB/s per edge, additional UCIe-S modules increase the required die area, and CoWoS-R becomes more economical over the remainder of the evaluated range.

An additional advantage of adopting standard packaging is more flexible die placement, guaranteeing a rectangular die design with an off-center UCIe interface. This flexibility is particularly important for CASCADE because advanced packaging imposes stricter physical constraints on die placement to avoid warpage problems and satisfy D2D reach constraints. Under these constraints, CASCADE cannot meet the advanced-packaging requirements, whereas standard packaging allows multiple placement options. For example, the placement in Figure~\ref{fig:sip_dse}(a) leverages the long channel reach of UCIe-S~\cite{ucie_3_0}, whereas the placement in Figure~\ref{fig:sip_dse}(b) leverages the less restrictive die-placement constraints of standard packaging. Both placements require nearly the same package size. We do not select between these placements because determining the optimal placement requires an in-depth mechanical analysis.

\subsubsection{Manufacturing Cost per Shipped SiP}
To evaluate the cost advantage of CASCADE's packaging-aware and homogeneous chiplet design, we compare the manufacturing cost per shipped SiP for CASCADE and AMMA. Because the AMMA-Compute chiplet lacks an external I/O interface, we assume that an AMMA SiP consists of 16 AMMA-Compute chiplets and one additional AMMA-I/O chiplet with a 1500~GB/s UCIe interface and a 16-lane PCIe interface. We exclude DRAM core die costs from both designs to focus the comparison on logic-die fabrication and packaging costs.

At a volume of 100{,}000 shipped SiPs, CASCADE has a 38.4\% lower manufacturing cost per shipped SiP than AMMA (\$1{,}897 versus \$3{,}081). As shown in Figure~\ref{fig:sip_cost}(b), CASCADE is more cost-efficient than AMMA across all evaluated production volumes. The cost breakdown in Figure~\ref{fig:sip_cost}(c) explains the sources of this cost gap. AMMA inevitably requires expensive advanced packaging to support its high D2D bandwidth, and its heterogeneous chiplet design incurs higher NRE costs. Furthermore, AMMA's inefficient die dimensions increase the RE cost of the 4$\times$4 tile.

\begin{figure*}
\centering
 \includegraphics[width=1.0\linewidth]{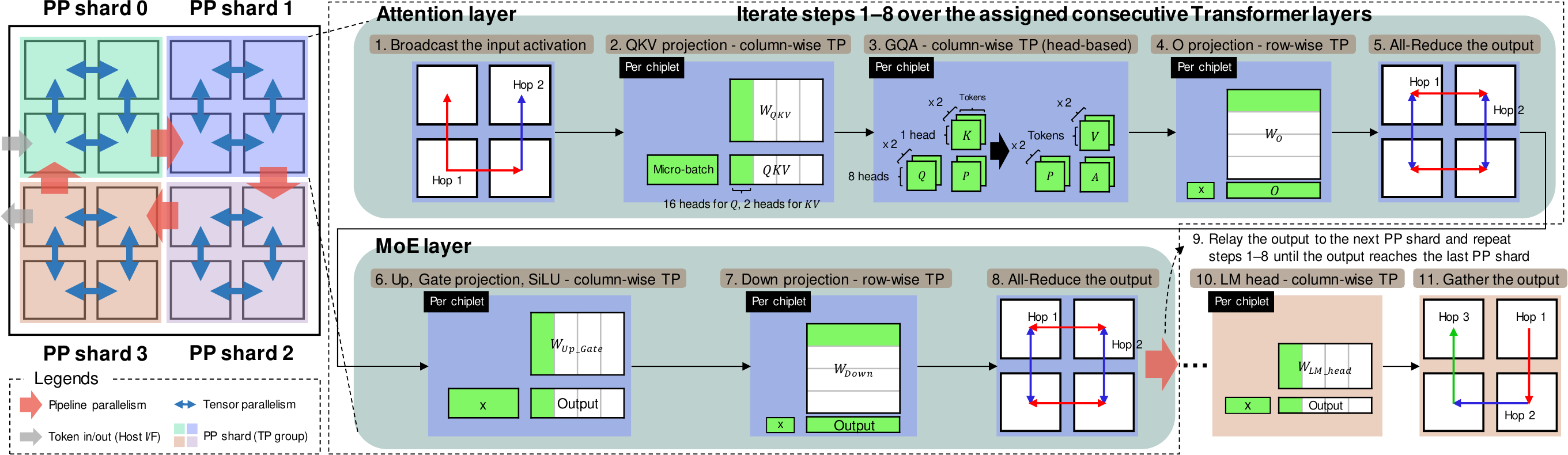}
\caption{TP+PP execution in CASCADE, using GPT-OSS-120B with a micro-batch size of 1 per PP shard as an example. Adjacent 2$\times$2 chiplets form a PP shard operating as a TP group. Activations are relayed between PP shards, and the final shard predicts tokens. Multi-device scaling increases the number of PP shards proportionally.}
\label{fig:cascade_2d_tp-pp}
\end{figure*}

\section{Parallel Processing Strategy}\label{sec:5}
CASCADE's cost-aware design matches compute throughput to the needs of memory-bound decoding while limiting D2D bandwidth and connectivity, making parallel execution critical to hardware utilization. We propose two forms of 2D parallelism whose relative performance varies across workloads, motivating the workload-specific DSE presented in Section~\ref{sec:7}.

\subsection{Limitations of One-Dimensional Parallelism}\label{sec:5.1}
In CASCADE, TP allows chiplets to load partitioned weights from local cHBMs in parallel but requires collectives at every decoder layer. Head-wise TP also faces a scalability limit in GQA. When the TP degree exceeds the KV head count, query heads are distributed across more chiplets while the corresponding KV heads are replicated~\cite{vllm_kv_replication}. This increases KV cache capacity requirements and redundant memory accesses, limiting further attention speedup.

Alternatively, partitioning each head's dimension across chiplets avoids KV replication. However, exact attention requires reducing partial QK scores before softmax~\cite{tpla}. This introduces an additional All-Reduce collective at each attention layer, incurring substantial communication overhead as its communication volume grows with context length~\cite{amma}. This makes head-dimension partitioning difficult to employ with CASCADE's limited D2D bandwidth.

PP assigns consecutive layers to separate chiplets and pipelines micro-batches across them. Micro-batching lowers the arithmetic intensity of linear projections, but each chiplet must load the weights of its assigned layers. Therefore, unlike TP, PP-only does not distribute a layer's weight accesses across chiplets. When the batch is too small for all chiplets to process micro-batches concurrently, some pipeline stages remain idle, reducing chiplet utilization.

We exclude expert parallelism (EP) from our design space. EP distributes experts across chiplets, requiring All-to-All for dispatch and combine~\cite{deepspeed_moe}. Input-dependent routing can introduce load imbalance and execution time variation, causing synchronization stalls and reducing chiplet utilization~\cite{llep}. These costs are undesirable for CASCADE's small-batch decoding and bandwidth-constrained interconnects. Therefore, we use TP for MoE computation.

\subsection{Two-Dimensional Parallelism}\label{sec:5.2}
We combine complementary forms of parallelism to better utilize CASCADE's resources, balancing the computation, memory access, and communication costs of one-dimensional execution.

\subsubsection{TP+PP}\label{sec:5.2.1}
Figure~\ref{fig:cascade_2d_tp-pp} shows the execution flow of TP+PP. The 16 cHBM chiplets are partitioned into four PP shards, each comprising a $2\times2$ cluster of adjacent chiplets. Each shard operates as a four-chiplet TP group and processes consecutive Transformer layers. For batch sizes of four or larger, we divide each batch into four micro-batches and pipeline them across the shards.

Within each TP group, QKV projections use column-wise TP, allowing each chiplet to process its assigned GQA heads locally. MoE gate and up projections also use column-wise TP, followed by local element-wise activations. The O and down projections use row-wise TP followed by All-Reduce. Activations are relayed between PP shards until the final shard executes the LM head. This confines layer-wise collectives to four-chiplet groups and inter-group transfers to pipeline stage boundaries. Moreover, despite the three-directional D2D interface limit per chiplet, TP+PP confines active communication to at most three directions, complying with the connectivity constraints of the homogeneous chiplet architecture.

\begin{figure}
\centering
 \includegraphics[width=1.0\linewidth]{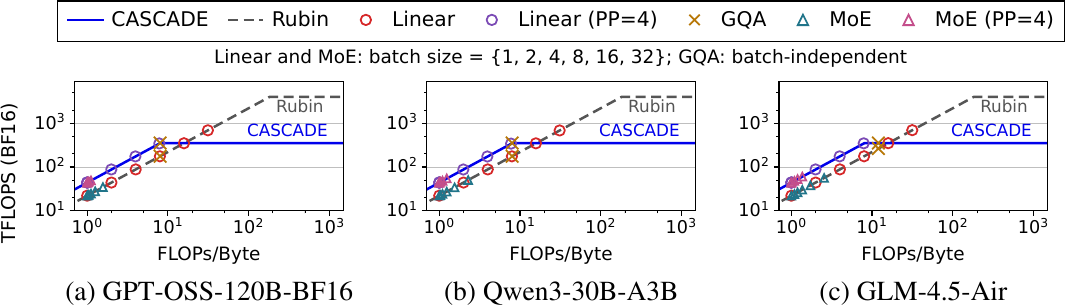}
\caption{Roofline analysis of CASCADE and Rubin for BF16 decoding across batch sizes 1{\textendash}32. PP=4 denotes four-way micro-batching under CASCADE's 2D TP+PP execution.}
\label{fig:roofline}
\end{figure}

Figure~\ref{fig:roofline} illustrates the roofline characteristics. TP+PP combines distributed weight accesses within each shard with smaller micro-batches across pipeline stages. MoE grouped GEMMs retain low arithmetic intensity, while GQA's arithmetic intensity remains approximately batch-independent. At batch size 32, four-way micro-batching reduces the arithmetic intensity of linear projections from approximately 32 to 8 FLOPs/Byte, approaching CASCADE's ridge point. This shifts linear projections toward the memory-bound regime, allowing them to better exploit its high memory bandwidth.

\begin{figure*}
\centering
 \includegraphics[width=1.0\linewidth]{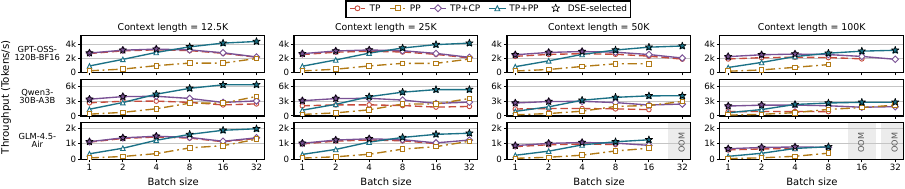}
\caption{Parallelism DSE across batch sizes and context lengths. Missing points indicate OOM.}
\label{fig:results_parallelism_dse}
\end{figure*}

\subsubsection{TP+CP}\label{sec:5.2.2}
We adopt TP+CP based on vLLM's decode context parallelism for GQA models such as Qwen3~\cite{vllm_dcp}. KV heads are assigned to separate chiplet groups, with each head's KV cache partitioned along the sequence dimension within its group. For GQA, we use TP=4 and CP=4 for Qwen3-30B-A3B, and TP=8 and CP=2 for GPT-OSS-120B-BF16 and GLM-4.5-Air.

Parallelism varies by operator. Q and O projections and MoE grouped GEMMs use TP across all 16 chiplets. K and V projection weights are partitioned by KV head across groups and replicated within each group. Within each CP group, all chiplets compute the new token's KV, but only the chiplet assigned to that token's sequence position stores it in the KV cache. Interleaved placement assigns tokens across the group without cache replication~\cite{vllm_kv_replication,vllm_dcp}.

After Q projection, an intra-group All-Gather makes the associated query heads available on every chiplet. Each chiplet computes partial attention over its local KV segment. A Reduce-Scatter jointly merges partial outputs and softmax statistics and distributes the results to their query-head owners. Finally, the row-wise O projection is followed by a package-wide All-Reduce.

Compared with KV-replicating TP, TP+CP avoids KV cache replication and redundant reads. Moreover, its additional communication volume is independent of context length because CP exchanges only queries and partial attention states. These benefits can outweigh the added communication cost for KV-intensive workloads.

\section{Methodology}\label{sec:6}
\subsection{Simulation Methodology}\label{sec:6.1}
We evaluate all accelerators using a unified in-house simulation framework implemented in Python. Computation costs are calculated from operation counts and the tensor and SIMD throughputs. Memory access and communication costs are determined from data volumes, bandwidths, and transfer latencies. The framework includes a command-level, event-driven model of chiplet execution. This model translates workload operations into computation and data-transfer commands according to the selected parallelism.

The scheduler decides when commands execute based on data dependencies and resource availability. It models double-buffered overlap between computation and HBM accesses across tiles, communication overlap, and shared-resource contention. D2D transfers follow chiplet topology and contend for links along their routes.

\begin{table}[t]
\centering
\caption{Per-package hardware specifications.}
\label{tab:hardware_specs}
\footnotesize
\setlength{\tabcolsep}{3pt}
\resizebox{\columnwidth}{!}{%
\begin{tabular}{lrrrrr}
\toprule
Accelerator & BF16 TFLOPS & HBM BW (TB/s) &
Cap. (GB) & Inter-chip (GB/s) & TDP (W) \\
\midrule
Rubin      & 4,000 & 22 & 288 &
1,800\ \makebox[2.8em][l]{(C2C)} & 2,300 \\
AMMA       &   768 & 44 & 576 &
1,500\ \makebox[2.8em][l]{(D2D)} & 1,440 \\
Duplex PIM & 170.4 & 88 & 288 &
2,750\ \makebox[2.8em][l]{(HBM)} & Unknown \\
CASCADE    &   352 & 44 & 384 &
128\ \makebox[2.8em][l]{(D2D)} & 1,052 \\
\bottomrule
\end{tabular}
}

\smallskip
\begin{minipage}{\columnwidth}
\scriptsize
Inter-chip bandwidth is per direction.
For Duplex PIM, it denotes the per-stack HBM bandwidth to the xPU.
All accelerators use 128 GB/s PCIe bandwidth per direction.
AMMA and Duplex PIM are assumed to have the same FP32 SIMD
throughput as CASCADE (4,096 GFLOPS per package),
while Rubin's is 130 TFLOPS.
Duplex PIM shares its HBM capacity with the xPU.
\end{minipage}
\end{table}

\subsection{Evaluation Setup}\label{sec:6.2}
Table~\ref{tab:hardware_specs} lists the hardware specifications. We compare CASCADE with state-of-the-art designs: Rubin~\cite{rubin}, AMMA+Rubin~\cite{amma}, and Duplex~\cite{duplex}. AMMA+Rubin splits devices equally, with AMMA executing attention and Rubin the remaining operations. For a fair comparison, we apply Rubin and HBM4 specifications to Duplex.

We evaluate GPT-OSS-120B-BF16~\cite{gptoss}, Qwen3-30B-A3B~\cite{qwen3}, and GLM-4.5-Air~\cite{glm45}, with weights, activations, and KV caches stored in BF16. All accelerators use identical synthetic expert-routing traces that maximize the number of distinct experts activated within a batch. All experiments except workload sensitivity use 1024 input tokens and their specified context lengths. Workload sensitivity uses the specified input and final sequence lengths. For CASCADE, we explore TP, PP, TP+PP, and TP+CP for each workload and select the one maximizing throughput under per-chiplet memory capacity constraints. If none is feasible, we report out-of-memory (OOM).

Throughput is batch size divided by the interval between consecutive decode step completions during continuous decoding, accounting for inter-iteration overlap. Energy efficiency (Tokens/J) is throughput divided by the total TDP of participating accelerators.

\subsection{Power Estimation}\label{sec:6.3}
CASCADE's PEs are modeled as systolic arrays. Their power is estimated using normalized energy efficiency (TOPS/W) based on state-of-the-art NPU~\cite{rebel_isscc26}. At a target BF16 throughput of 22 TFLOPS per cHBM, the combined PE power is 9 W. For RTL-based digital logic power estimation, we use Synopsys RTL Architect~\cite{rtl_architect}. Physically-aware analysis follows synthesis, placement and routing, and gate-level matching. The analysis is vectorless, assuming a toggle ratio of 0.2 at the TT corner, 0.75 V, and 85$^\circ$C.

HBM core die power is evaluated using Cadence Sigrity X~\cite{sigrity_x} under JEDEC IDD4R and IDD4W patterns~\cite{jedec_hbm4} for reads and writes, respectively, assuming a half-toggle condition with 50\% DQ switching activity. We calculate the weighted average power over active read/write bursts, excluding idle and state-transition intervals.

\section{Evaluation Results}\label{sec:7}

\subsection{Parallelism Design Space Exploration}\label{sec:7.1}
Figure~\ref{fig:results_parallelism_dse} compares TP, PP, TP+PP, and TP+CP across batch sizes and context lengths. Among the 69 feasible configurations, TP+CP achieves the highest throughput in 32 and TP+PP in 37. Neither TP-only nor PP-only is selected, demonstrating the benefits of combining complementary forms of parallelism on CASCADE.

At very small batch sizes, TP+CP benefits from package-wide parallel weight accesses without the pipeline underutilization or repeated weight reads from micro-batching. As batch size grows, TP+PP improves pipeline utilization and limits the growth of TP collective communication overhead. This tradeoff determines when TP+PP outperforms TP+CP. In TP+PP, communication-induced pipeline bubbles are negligible because the next micro-batch arrives before the bottleneck stage finishes the current one. Each inter-stage transfer takes only 0.006{\textendash}0.026\% of the steady-state period. Subsequent experiments use each workload's optimal strategy.

\begin{figure}
\centering
 \includegraphics[width=1.0\linewidth]{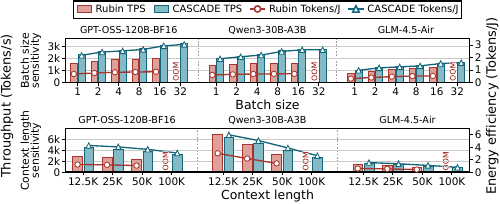}
\caption{Single-device evaluation results. Batch size sensitivity uses context lengths 100K/100K/25K; context length sensitivity uses batch sizes 32/32/8, in model order.}
\label{fig:results_single_device}
\end{figure}

\begin{figure}
\centering
 \includegraphics[width=1.0\linewidth]{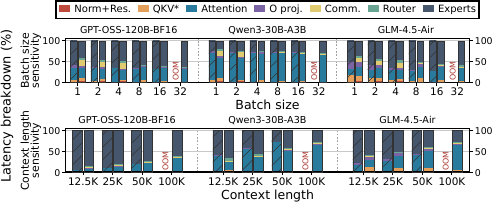}
\caption{Per-layer latency breakdown for Rubin (hatched) and CASCADE (solid), with each bar normalized to 100\%. LM head and host I/O are excluded. QKV* includes RoPE and KV append. For GLM-4.5-Air, Experts includes the dense MLP.}
\label{fig:results_latency_breakdown}
\end{figure}

\subsection{Single-Device Evaluation}\label{sec:7.2}
Figure~\ref{fig:results_single_device} compares throughput and energy efficiency of CASCADE and Rubin. CASCADE achieves up to $1.71\times$ higher throughput and $3.74\times$ higher energy efficiency. Its higher memory bandwidth benefits memory-intensive decoding, while its lower power improves energy efficiency. CASCADE's 384 GB capacity supports all plotted workloads, whereas Rubin's 288 GB results in OOM for some large-batch or long-context workloads. For Qwen3-30B-A3B at batch size 32 and a 12.5K context, CASCADE delivers lower throughput than Rubin. Limited KV traffic and reduced expert weight reuse under micro-batching limit its bandwidth advantage. Nevertheless, CASCADE maintains approximately $1.99\times$ Rubin's energy efficiency.

Figure~\ref{fig:results_latency_breakdown} shows a substantial drop in CASCADE's communication share when switching from TP+CP to TP+PP. This highlights the importance of workload-driven DSE, as the optimal 2D strategy under hardware resource constraints varies across workloads.

\begin{figure}
\centering
 \includegraphics[width=1.0\linewidth]{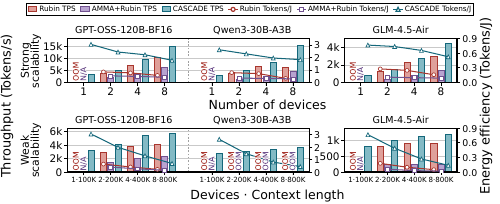}
\caption{Scalability evaluation at fixed batch sizes 32/32/8 in model order, with a fixed 100K context for strong scaling.}
\label{fig:results_scalability}
\end{figure}

\subsection{Scalability}\label{sec:7.3}
Figure~\ref{fig:results_scalability} presents strong and weak scalability results. Under strong scaling with eight devices, CASCADE achieves up to $2.52\times$ Rubin's throughput and $3.51\times$ that of AMMA+Rubin, with higher energy efficiency. In AMMA+Rubin, memory-bound MoE layers execute on Rubin, whose lower HBM bandwidth increases MoE execution latency compared to CASCADE. AMMA+Rubin also transfers activations between accelerators at every Transformer layer, whereas CASCADE communicates only at inter-device PP stage boundaries.

AMMA's reported 32 KB SRAM per core limits KV blocks to 8{\textendash}32 tokens, whereas CASCADE's 2 MiB per PE supports 128-token blocks. Smaller blocks require more online softmax updates, making attention SIMD-bound. With one AMMA and one Rubin, GLM-4.5-Air's slowest attention layer (batch 4, 25K context) takes $8.4\times$ the HBM-bandwidth-limited time. This SIMD bottleneck limits gains from AMMA's high tensor throughput and D2D bandwidth.

Under weak scaling, context length increases by 100K per device while batch size remains fixed. CASCADE supports all models up to an 800K context, whereas Rubin and AMMA+Rubin encounter OOM for Qwen3-30B-A3B. All platforms exhibit decreasing energy efficiency as power grows faster than throughput, while CASCADE remains more energy-efficient than Rubin and AMMA+Rubin.

Table~\ref{tab:results_packaging_scalability} compares CASCADE's $4\times4$ and $6\times6$ packages. The latter uses nine four-chiplet TP groups as pipeline stages while maintaining the three-directional D2D constraint. Expanding from 16 to 36 chiplets increases memory capacity and improves throughput by $1.60\times${\textendash}$1.98\times$, but reduces energy efficiency to 71{\textendash}88\% of the $4\times4$.

\begin{table}[t]
\centering
\caption{Package-level scalability at a context length of 100K.}
\label{tab:results_packaging_scalability}
\footnotesize
\setlength{\tabcolsep}{1.5pt}
\setlength{\cmidrulekern}{1pt}
\begin{tabular*}{\columnwidth}{@{\extracolsep{\fill}}lrrrrrr@{}}
\toprule
\multicolumn{1}{c}{
  \raisebox{-1.5ex}[0pt][0pt]{Model (Batch size)}
}
& \multicolumn{2}{c}{Latency (ms)}
& \multicolumn{2}{c}{Tokens/s}
& \multicolumn{2}{c}{Tokens/J} \\
\cmidrule(lr){2-3}\cmidrule(lr){4-5}\cmidrule(lr){6-7}
& $4\times4$ & $6\times6$
& $4\times4$ & $6\times6$
& $4\times4$ & $6\times6$ \\
\midrule
GPT-OSS-120B-BF16 (32)
& 9.975 & \textbf{5.026}
& 3,208 & \textbf{6,366}
& \textbf{3.049} & 2.689 \\
Qwen3-30B-A3B (32)
& 11.617 & \textbf{6.087}
& 2,755 & \textbf{5,257}
& \textbf{2.618} & 2.221 \\
GLM-4.5-Air (8)
& 9.898 & \textbf{6.189}
& 808 & \textbf{1,293}
& \textbf{0.768} & 0.546 \\
\bottomrule
\end{tabular*}
\end{table}

\subsection{Power Analysis and Ablation Study}\label{sec:7.4}
\subsubsection{Power Analysis}\label{sec:7.4.1}
We compare power per cHBM chiplet. At 32 GT/s and 0.7 V, the UCIe energy-per-bit targets are $E_S=1.25$ pJ/bit for UCIe-S and $E_A=0.6$ pJ/bit for UCIe-A~\cite{ucie_3_0}. For respective bandwidths $B_S$ and $B_A$, we compare their D2D power as $P_S = P_A (B_S/B_A)(E_S/E_A)$. At equal bandwidth, $P_S/P_A=1.25/0.6\approx2.08$. Thus, standard packaging requires lower D2D bandwidth under a fixed power budget. Using these targets, 128 GB/s with UCIe-S and 256 GB/s with UCIe-A require comparable D2D power.

We compare CASCADE with a variant using advanced packaging, 256 GB/s D2D bandwidth, and 12-Hi stacks. The variant doubles D2D bandwidth at comparable D2D power. Including additional core-die power, chiplet power rises from 65.76 W to 67.43 W (2.5\%).

We estimate AMMA's power using its compute throughput, D2D bandwidth, and number of D2D interfaces in CASCADE's power model. The estimated 93.21 W is 3.6\% above AMMA's reported 90 W~\cite{amma}. AMMA's estimated PE and D2D power are respectively $2.18\times$ and $3.75\times$ CASCADE's. These increases explain most of the power difference, highlighting the power cost of higher compute throughput and D2D bandwidth despite limited decoding speedups.


\begin{table}[t]
\centering
\caption{CASCADE ablation study in Tokens/J.}
\label{tab:results_ablation}
\footnotesize

\renewcommand{\arraystretch}{1.0}
\setlength{\tabcolsep}{1pt}
\setlength{\aboverulesep}{0.4ex}
\setlength{\belowrulesep}{0.65ex}
\setlength{\cmidrulekern}{1pt}

\newlength{\gptShift}
\newlength{\gptAbsWidth}
\newlength{\gptRatioWidth}
\settowidth{\gptShift}{GPT-OSS-120B-BF16 (100K)}
\settowidth{\gptAbsWidth}{$\alpha$ (abs.)}
\settowidth{\gptRatioWidth}{$0.99\times$}
\setlength{\gptShift}{%
  \dimexpr(\gptShift-2\gptAbsWidth
  -\gptRatioWidth-4\tabcolsep)/2\relax}
\newcommand{\gptMove}[1]{%
  \hspace{\gptShift}#1\hspace{-\gptShift}}

\newlength{\abBatchWidth}
\settowidth{\abBatchWidth}{32}
\newcommand{\abBatch}[1]{%
  \makebox[\abBatchWidth][r]{#1}}

\resizebox{\columnwidth}{!}{%
\begin{tabular}{@{}*{10}{c}@{}}
\toprule
\raisebox{-1.5ex}[0pt][0pt]{Batch}
& \multicolumn{3}{c}{GPT-OSS-120B-BF16 (100K)}
& \multicolumn{3}{c}{Qwen3-30B-A3B (100K)}
& \multicolumn{3}{c}{GLM-4.5-Air (25K)} \\
\cmidrule(lr){2-4}
\cmidrule(lr){5-7}
\cmidrule(lr){8-10}
& \gptMove{$\alpha$ (abs.)}
& \gptMove{$\beta/\alpha$}
& $\gamma/\beta$
& $\alpha$ (abs.) & $\beta/\alpha$ & $\gamma/\beta$
& $\alpha$ (abs.) & $\beta/\alpha$ & $\gamma/\beta$ \\
\midrule
\abBatch{1}
& \gptMove{2.214} & \gptMove{$0.99\times$} & $1.00\times$
& 1.913 & $0.99\times$ & $1.00\times$
& 0.986 & $0.99\times$ & $1.00\times$ \\
\abBatch{2}
& \gptMove{2.484} & \gptMove{$0.98\times$} & $1.00\times$
& 2.084 & $0.99\times$ & $1.00\times$
& 1.188 & $0.99\times$ & $1.00\times$ \\
\abBatch{4}
& \gptMove{2.585} & \gptMove{$0.98\times$} & $1.00\times$
& 2.104 & $0.99\times$ & $1.06\times$
& 1.270 & $0.98\times$ & $1.00\times$ \\
\abBatch{8}
& \gptMove{2.511} & \gptMove{$0.98\times$} & $1.08\times$
& 1.966 & $0.99\times$ & $1.27\times$
& 1.210 & $0.99\times$ & $1.12\times$ \\
\abBatch{16}
& \gptMove{2.248} & \gptMove{$0.98\times$} & $1.33\times$
& 1.672 & $1.00\times$ & $1.57\times$
& 1.010 & $0.99\times$ & $1.52\times$ \\
\abBatch{32}
& \gptMove{1.840} & \gptMove{$0.99\times$} & $1.68\times$
& 1.799 & $0.99\times$ & $1.47\times$
& 1.192 & $0.98\times$ & $1.35\times$ \\
\bottomrule
\end{tabular}%
}

\begin{minipage}{\columnwidth}
\scriptsize
\raggedright
$\alpha$: A-256-12Hi;
$\beta$: S-128-8Hi;
$\gamma$: $\beta$+DSE (\textbf{CASCADE}).
Both $\alpha$ and $\beta$ use only TP+CP.
A and S denote advanced and standard packaging, respectively;
numbers specify D2D BW (GB/s) and HBM height.
\end{minipage}
\end{table}

\begin{figure}
\centering
 \includegraphics[width=1.0\linewidth]{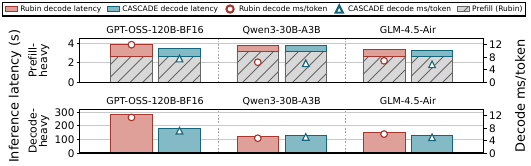}
\caption{Workload sensitivity at batch sizes 32/32/8 in model order. Input/output sequence lengths are 25K/100 for prefill-heavy and 1024/25K for decode-heavy.}
\label{fig:results_workload_sensitivity}
\end{figure}

\subsubsection{Ablation Study}\label{sec:7.4.2}
Table~\ref{tab:results_ablation} separates the effects of hardware and parallelism selection. With TP+CP fixed, $\beta$ uses standard packaging, half the D2D bandwidth, and fewer core dies while retaining approximately 98{\textendash}100\% of $\alpha$'s energy efficiency. This supports our hardware specification DSE, as standard packaging incurs only a small energy efficiency loss. This also suggests that 128 GB/s D2D bandwidth is sufficient for the evaluated workloads.

With the same hardware and power as $\beta$, $\gamma$ selects the optimal parallelism through DSE. The resulting throughput gains improve energy efficiency by up to $1.68\times$. At larger batch sizes, $\gamma$ achieves higher energy efficiency than $\alpha$, demonstrating the benefit of parallelism selection on the standard-packaged design.

\subsection{Workload Sensitivity}\label{sec:7.5}
Figure~\ref{fig:results_workload_sensitivity} shows inference latency and average per-token decode time. We use relatively short final sequence lengths to assess CASCADE's sensitivity to potentially unfavorable conditions with limited KV traffic. Prefill runs on Rubin for both accelerators. In prefill-heavy workloads, prefill dominates latency in both systems because decoding requires few steps. This limits the end-to-end latency reduction from CASCADE's faster decoding to at most 11.3\%.

For decode-heavy workloads, CASCADE reduces inference latency by up to 37.1\%. However, CASCADE's latency is 8.3\% higher than Rubin's for Qwen3-30B-A3B because Rubin executes many early short-context decode steps faster, consistent with Section~\ref{sec:7.2}.

\begin{figure}
\centering
 \includegraphics[width=1.0\linewidth]{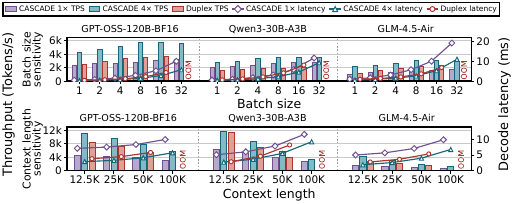}
\caption{CASCADE bandwidth scaling compared with Duplex. CASCADE $1\times$ and $4\times$ use per-stack internal bandwidths of 2.75 and 11 TB/s, respectively; Duplex uses 11 TB/s.}
\label{fig:results_bandwidth_scaling}
\end{figure}

\subsection{Memory Bandwidth Scaling}\label{sec:7.6}
Figure~\ref{fig:results_bandwidth_scaling} compares CASCADE with Duplex, including a CASCADE variant with fourfold per-stack internal bandwidth matching Duplex's 11 TB/s. CASCADE $4\times$ achieves up to $2.64\times$ the throughput of its nominal design and $2.00\times$ that of Duplex. This bandwidth expansion is an orthogonal design option that complements CASCADE's array-based bandwidth scaling.

However, increasing bandwidth alone lowers the ridge point and can make memory-bound operations compute-bound. With HBM4, Duplex executes MoE on Logic-PIM, while GQA runs on the xPU because Logic-PIM's ridge point of approximately 1.94 FLOPs/Byte makes GQA compute-bound, limiting its roofline throughput to 170.4 TFLOPS. In comparison, the xPU's memory-bound roofline estimates are 176 and 264 TFLOPS for GQA at 8 and 12 FLOPs/Byte, respectively. Thus, exploiting this bandwidth requires higher compute throughput, straining the logic die's area and power budgets.

\section{Conclusion}\label{sec:8}
This paper presented CASCADE, a cost-aware memory-centric chiplet-array accelerator for MoE-LLM decode. Through workload-driven DSE, CASCADE co-designs hardware specifications and parallel execution strategies while considering physical feasibility and manufacturing cost. By leveraging compute-enabled cHBM chiplets and workload-specific parallelism, CASCADE achieves higher performance and energy efficiency than state-of-the-art accelerators with cost-effective standard organic-substrate packaging. These results demonstrate the importance of hardware--software co-design for practical memory-centric LLM decoding accelerators.


\bibliographystyle{ACM-Reference-Format}
\bibliography{refs}

\end{document}